\documentclass[final,5p,10pt,times,twocolumn,year]{elsarticle}

\usepackage{amssymb}
\usepackage{amsmath}
\usepackage{slashed}
\usepackage{mathrsfs}
\usepackage{url}

\journal{---}

\begin{document}
\begin{frontmatter}



\title{On the Topological Nature of the Butterfly Effect}
\author[label1]{Igor V. Ovchinnikov}
\affiliation[label1]{organization={R\&D, CSD, ThermoFisher Scientific Inc},
            addressline={200 Oyster Point}, 
            city={South San Francisco},
            postcode={94080}, 
            state={CA},
            country={USA}}
\begin{abstract}
Non-integrability in the sense of dynamical systems, also known as dynamical chaos, is a strongly nonlinear qualitative phenomenon. Its most promising theoretical descriptions are likely to emerge from non-perturbative approaches, with symmetry-based methods being particularly reliable. One such symmetry-based framework is supersymmetric theory of stochastic dynamics (STS). STS reformulates a general form stochastic (partial) differential equations (SDE) as a cohomological topological field theory (TFT) and identifies the order associated with the spontaneous breakdown of the corresponding topological supersymmetry (TS) as the stochastic generalization of chaos. The Faddeev-Popov ghosts of STS act as a systematic bookkeeping tool for the dynamic differentials from the definition of the butterfly effect (BE): the infinitely long dynamical memory unique to chaos. Accordingly, the effective field theory (EFT) of the TS breaking is essentially a field theory of the BE in the long-wavelength limit. Building on this perspective, here we demonstrate that one way to build such EFTs is the background field method with the external $\mathfrak{gl}(1|1)$ supergauge field coupled to N=2 supercurrents of STS, the fermion number conservation, and translations in time. By the Goldstone theorem, the resulting EFTs are conformal field theories (CFTs) and the operator product expansion provides an explanation for 1/f noise -- the experimental signature of chaos in the form of dynamical power-law correlations. Moreover, the generating functional of the background field possesses its own TS, revealing the topological nature of the BE. Particularly, when the Anti-de Sitter(AdS)/CFT correspondence is an acceptable approximation, the holographic dual of EFT is a cohomological TFT on AdS in which the associated TS and the isometry of the basespace underlie the BE and 1/f noise, respectively.
\end{abstract}
\begin{keyword}
Topology \sep Supersymmetry \sep Symmetry Breaking \sep Chaos \sep Topological Field Theory
\end{keyword}
\end{frontmatter}

\section{Introduction}
The butterfly effect (BE) is a defining characteristic of the non-linear dynamical phenomenon commonly referred to as chaos \cite{Mot14,Rue14,Poincare_celestial_dynamics,ButterFly}. It is traditionally described as an infinitely persistent dynamical memory, wherein even a minor perturbation alters the trajectory forever and, consequently, is never forgotten. As to chaos itself, the modern paradigm views it as a disordered state, with the BE serving as mere evidence of the dynamical instability and unpredictability. Once the presence of the BE is established, the dynamics is classified as chaotic, and the role of the BE is considered fulfilled. Deeper exploration of the BE is rarely undertaken, which is understandable: despite many important results in chaos theory (see, e.g., Refs. \cite{Gil98,Gilmore_book,RevModPhys.57.617,Chaos_book_1,Chaos_book_2017,10.1063/5.0025924,Yorke_1975,TransientChaos2015,Intermittency_Review,Baxendale.10.1007/BFb0076851,Arnold.10.1007/BFb0076835} and Refs. therein), a consistent theory that, for instance, addresses the spatiotemporal long-range structure of the BE has yet to emerge. 

Such a theory, however, could become a valuable addition to chaos theory, benefit a range of disciplines,
\footnote{In neurodynamics, for instance, there are reasons to believe that the BE plays an important role in the short-term memory and consciousness \cite{ovchinnikov2020}.}
and shed new light on other unresolved chaos-related problems such as rigorous explanation of 1/f noise -- the experimental manifestation of scale invariance of chaos observed universally across various fields, including physics \cite{Voss_1979,RevModPhys.53.497,Keshner_1_f_noise_1982,Wang_1fnoise}, biology \cite{pinkNoiseBiosystems2001,EEG,Biology1fNoise,BookHeartBrainNoise}, econodynamics \cite{Pre11}, neurodynamics \cite{PhysRevLett.97.118102} \emph{etc.}


The direction to look for a theory with such multidisciplinary applicability is stochastic (partial) differential equations (SDEs) \cite{Oks10,Baxendale1,Watanabe1,Crauel1,Baxendale.10.1007/BFb0076851,Arnold.10.1007/BFb0076835, ARNOLD19831,Oks10,LEJAN1984307,Kunita2019,Hairer_2001,kupiainen2016renormalization,Heirer_2018,Bedrossian_RecentReview}, the class of models of unparalleled relevance in modern science. SDEs cover virtually all scientific disciplines outside of physics, while within physics, they cover and extend beyond classical physics.\footnote{In classical physics, only a small part of models -- those where energy dissipation can be neglected -- belongs to the domain of classical mechanics, Hamilton dynamics, and least action principles. In more general cases, equations of motion (EoM) are "nonconservative" and have no underlying least action principle. In even more general cases, the very concept of energy becomes irrelevant, as in Lorenz model. \cite{Lorenz} Therefore, a general \emph{deterministic} classical physics model belongs to the domain of dynamical systems theory (see, e.g., Refs.\cite{Arn03,Gil98,Teschl,Izhikevich,Chaos_book_1,Chaos_book_2} and Refs. therein), \emph{i.e.}, a generalization of Hamilton dynamics to EoMs of arbitrary forms. Furthermore, all \emph{natural} systems are influenced by random noise from their environments. That is, a most general classical physics model is described by an SDE of a general form.} 
Moreover, one recent approach to SDEs called the supersymmetric theory of SDEs (STS, see, \emph{e.g}, Ref.\cite{OVCHINNIKOV2024114611} and Refs. therein) has introduced a symmetry-based framework showing that chaos, contrary to the semantics of this term and its traditional perception, is a spontaneous long-range order. This fact alone implies that the BE must possess a nontrivial spatiotemporal long-range structure, which may be worth deeper exploration.

STS builds on the Parisi-Soulas approach to Langevin SDEs \cite{ParSour,ParSour1,ZinnJustin}, or more precisely, its interpretation \cite{Baulieu_1988} as a cohomological topological field theory (TFT) \cite{TFT_BOOK,labastida1989,Baulieu_1988,Baulieu_1989,Witten_1982,Witten98,Witten981}, and extends it to SDEs of arbitrary form.\footnote{STS can also be viewed as an application of the generalized transfer operator approach of dynamical systems theory to SDEs \cite{Rue02}. It can also be looked upon as a stochastic counterpart of the deterministic topological theory of chaos \cite{Gilmore_book}.} This framework reformulates an SDE as a cohomological TFT and recognizes that the spontaneous breakdown of the topological supersymmetry (TS) inherent in this class of models as the stochastic generalization of dynamical chaos.

The key distinction between STS and the traditional Fokker-Planck framework for stochastic evolution lies in the structure of the wavefunctions -- the mathematical entities representing the dynamic state of an external observer. In STS, the  wavefunctions depend not only on the original variables of the SDEs but also on their supersymmetric partners known as Faddeev-Popov ghosts. These fields represent dynamic differentials that determine the BE. Accordingly, in Ref. \cite{OVCHINNIKOV2024114611} it was proposed that the effective field theory (EFT) for TS breaking in STS is essentially a consistent physical theory of the BE. 

Here, we address some general properties of these EFTs. We use an adaptation of the well-established concept of the background gauge-field from the theory of spontaneous breakdown of global bosonic symmetries \cite{RFT_SSB_book}. In case of STS, the key element of this approach is the background $\mathfrak{gl}(1|1)$ supergauge field coupled to conserved currents of STS. A particularly intriguing outcome of this approach is that the resulting generating functional of the supergauge field is not only supergauge invariant, as expected in any application of the background gauge field method, but it also possesses its own TS. This particularly suggests that in scenarios where the Anti-de-Sitter(AdS)/Conformal Field Theory(CFT) correspondence \cite{AdSCFT_Maldacena,GUBSER1998105,Witten1998AntideSS,natsuume2016adscftdualityuserguide,Penedones_2016,KaplanAdSCFT} is an acceptable approximation, the holographic dual of the EFT is a cohomological TFT on AdS.

The paper is organized as follows. After a brief introduction of the key elements of STS in Sec.\ref{Sec:STS}, we discuss the global symmetries of STS and introduce the external background $\mathfrak{gl}(1|1)$ supergauge field in Sec.\ref{Symmetries}. Sec.\ref{Sec:EFT} focuses on cenrtal features of EFTs, putting forward the idea that the generating functional must possess both the supergauge invariance and TS. In Sec.\ref{Sec:CFT}, we discuss the CFT aspects of EFTs and how the operator product expansion explains 1/f noise. Finally, in Sec.\ref{Generating_functional}, we show that the leading terms of the generating functional do possess TS and that in situations where the AdS/CFT correspondence is applicable, the CFT holographic dual of the EFT is a TFT.

\section{Stochastic Dynamics and Topological Supersymmetry}
\label{Sec:STS}
\subsection{Stochastic Evolution Operator}
Most of the models in the literature belong to the following general class of SDEs,
\begin{eqnarray}
    \partial_t \varphi(x) = f(\varphi,x) + g_a(\varphi,x)\xi^a(x) \equiv F,\label{SDE}
\end{eqnarray}
where $x=(t,r)$ with $t$ being time and $r$ being the point in a $(d-1)$-dimensional "spatial part" of the basespace, $\varphi$ is a field from some target space, which for the sake of concreteness can be though of as a compact, orientable, smooth manifold, $f$ is a vector field from the tangent space representing the deterministic law of evolution, $\xi$ is a set of noise variables, which is assumed Gaussian white, \emph{i.e.}, with the probability $P(\xi) \propto e^{-\int \xi(x)^2 d^{d}x/2}$, and $\langle\xi^a(x)\xi^b(x')\rangle=\delta^{ab}\delta^{d}(x-x')$, and $g$ is a set of vector fields that couple the noise to the system. Even though Eq.(\ref{SDE}) is given in a more general form, we will only speak of translationally invariant models. 

Stochastic quantization \cite{ZinnJustin,ParSour,ParSour1,Baulieu_1988,Baulieu_1989} can be interpreted as a procedure of the substitution of the noise variables by the dynamic variables of the model using Eq.(\ref{SDE}) as a relation between the two (see Ref.\cite{TFT_BOOK} for Langevin SDEs and Ref.\cite{OvcEntropy} for SDEs of general form):
\begin{eqnarray}
    W &=& \iint_{P.B.C}Det \frac{\delta(\partial_t \varphi - F)}{\delta \varphi} \delta(\partial_t \varphi - F) \mathcal{D}\varphi P(\xi) \mathcal{D}\xi \nonumber \\
    &=& \iint_{P.B.C} e^{-S(\Phi)} \mathcal{D}\Phi = \text{Tr } (-1)^{\hat N} \hat M_{t_ft_i}.\label{WittenIndex}
\end{eqnarray}
Here, the pathintegration goes over the closed paths as indicated by the subscript denoting periodic boundary conditions (P.B.C.). This is needed to make sure that the numbers of variables  of the noise and the system are the same so that the variable transformation is meaningful. The collection of fields, $\Phi=(\varphi,B,\bar\chi,\chi)$, contains the original field, $\varphi$, the Legandre multiplier, $B$, and a pair of Faddeev-Popov ghosts, $\bar\chi,\chi$, which are fermionic or Grassmann fields needed to account for the functional determinant in Eqs.(\ref{WittenIndex}) and (\ref{index}). The action,
\begin{eqnarray}
S = [\mathcal{Q}, \Psi], \label{Action_0}
\end{eqnarray}
where square brackets denote bi-graded (super)commutator, $[ a,b] = a b - (-1)^{\deg a \deg b} ba$, 
\begin{eqnarray}
    &&\mathcal{Q}= \int d^{d}x \left(\chi(x) \delta/{\delta \varphi(x)} + B(x) \delta/{\delta \bar\chi(x)}\right), \label{Eq_q}
\end{eqnarray}
is the pathintegral version of the TS, and\footnote{We drop indices unless this leads to a confusion.}
\begin{eqnarray}
    \Psi = \int  ( i\bar\chi \partial_t \varphi - \bar q ) d^{d}x,\; \bar q = i\bar\chi \left( f - i g L_{g}/2 \right), \label{barQ}
\end{eqnarray}
is the so-called gauge fermion with $L_{g} = [\mathcal{Q},i\bar\chi g]$ being the pathintegral version of Lie derivative along $g$. In Eq.(\ref{WittenIndex}), $\hat M$ and $\hat N$ are the stochastic evolution operator and the ghost/fermion number operator to be defined shortly and the alternating sign factor is due to the unconventional P.B.C. for fermions.

Eq.(\ref{WittenIndex}) is the famous Witten index. It can be represented as,
\begin{eqnarray}
        W &=& \iint_{P.B.C} Det \frac{\delta(\partial_t \varphi - F)}{\delta \varphi} \delta(\partial_t \varphi - F) \mathcal{D}\varphi P(\xi) \mathcal{D}\xi \nonumber \\ 
        &=& \iint P(\xi) \text{Ind}(\xi) \mathcal{D}\xi =  \text{Ind}\cdot Z_{noise}\label{WittenTopInv}
\end{eqnarray}
where $Z_{noise}=\iint P(\xi) \mathcal{D}\xi=1$ is the partition function of the noise and
\begin{eqnarray}
\text{Ind}(\xi)  = \sum_{closed\;solutions} sign Det \frac{\delta(\partial_t \varphi - F)}{\delta \varphi}= \text{Ind}\label{index},
\end{eqnarray}
is the index of the map from the space of the closed paths to the space of noise configurations.\footnote{The explicit form of the map is obtained by solving for $\xi$ in Eq.(\ref{SDE})} This index, along with the Witten index itself, is a topological invariant.\footnote{For example, in finite-dimentional model with compact phase space the index equals Euler characteristic of the phase space.} It does not depend on many things and particularly on the configuration of the noise, which is the reason why the last equality in Eq.(\ref{WittenTopInv}) is valid in the first place.

Being a topological invariant, the Witten index cannot be changed by any perturbation of SDE and all the "response correlators" vanish as discussed in \ref{WittenIndexButterflyEffect}. In other words, the Witten index cannot serve as the partition function of the model and this is not surprising -- as can be concluded from Eq.(\ref{WittenTopInv}), the Witten index (up to a topological factor) is the partition function of the noise and the noise has no information on the dynamics of the system whether perturbed or not. 

The actual partition function is obtained by switching to the anti-periodic conditions for fermionic fields,
\begin{eqnarray}
    Z = \iint_{A.P.B.C} e^{-S(\Phi)} \mathcal{D}\Phi = \text{Tr }\hat M_{t_ft_i},\label{partitionFunction}
\end{eqnarray}
where $\hat M_{t_ft_i}$ is the central element in the theory, the stochastic evolution operator (SEO). It can be expressed as follows,
\begin{eqnarray}
    &\langle \varphi\chi_f|\hat M_{t_ft_i} | \varphi\chi_i \rangle =  \iint_{\varphi\chi(t_{f,i}) =\varphi\chi_{f,i} }e^{-S(\Phi)} {\mathcal D}\Phi,\label{SEO} 
\end{eqnarray}
with $t_i$ and $t_f$ being the initial and final moments of evolution, the pathintegration goes over the open paths connecting the in- and out-arguments of the SEO, $|\varphi\chi\rangle$ and $\langle \varphi\chi|$ are the ket and the bra of the basis vectors of the operator representation where $\varphi$ and $\chi$ are diagonal so that $\hat \varphi(r)|\varphi\chi\rangle=\varphi(r)|\varphi\chi\rangle$, $\langle \varphi\chi|\hat \varphi(r)=\langle \varphi\chi| \varphi(r)$, (and the same for $\hat\chi$) and $\langle \varphi\chi_f|\varphi\chi_i \rangle = \prod_r\delta(\varphi_f(r)-\varphi_i(r))\delta(\chi_f(r)-\chi_i(r)) $, and the finite-time SEO,
\begin{eqnarray}
   \hat M_{t_ft_i} = e^{- (t_f-t_i)\hat H},
\end{eqnarray}
where the infinitesimal SEO,
\begin{eqnarray}
    \hat H = [\hat Q, \hat{\bar Q}],\label{SEOOper}
\end{eqnarray}
with
\begin{eqnarray}
    \hat Q = \int \chi(r) i\hat B(r) d^{d-1}r, \; \hat{\bar Q} = \int i\hat{\bar\chi} \left( f - i g \hat L_{f}/2 \right) d^{d-1} r,\label{Suerpcharges}
\end{eqnarray} being the operator versions of TS and operator $\bar q$ from Eq.(\ref{barQ}) obtained by the standard associations: $i\bar\chi(r) \to i\hat{\bar\chi}(r)=\delta/\delta\chi(r)$ and $iB(r)\to i\hat B(r)=\delta/\delta\varphi(r)$ corresponding to the operator representation where $\varphi$ and $\chi$ are diagonal.

It must be pointed out that the pathintegral representations of evolution operators like Eq.(\ref{SEO}) have an intrinsic ambiguity associated with the freedom of choosing the operator ordering convention on turning from pathintegrals to the operator representation. In the classical theory of SDEs, this ambiguity is known as the Ito-Stratonovich dilemma \cite{West}. In STS, this ambiguity can be removed by recognizing the most natural mathematical meaning of SEO which is that of the generalized transfer operator of dynamical systems theory \cite{Rue02} -- the pullback induced by the SDE-defined noise-specific diffeomorpisms of the phase space averaged over noise configurations. Such SEO is unique and it corresponds to the Stratonovich approach and/or the Weyl bi-graded symmetrization rule for operator ordering \cite{OvcEntropy}.

\begin{figure}[t] 
    \centering
\includegraphics[width=0.7\linewidth,height=3.0cm]{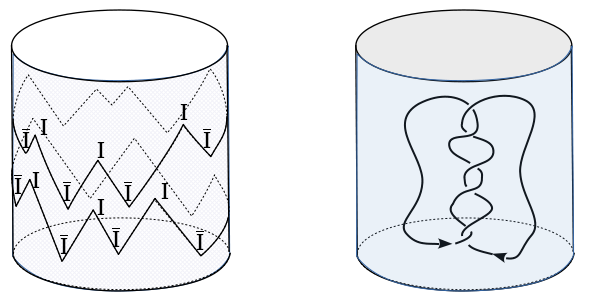}
\caption{\label{Fig_1} {\bf (left)} In the overdamped sine-Gordon equation \cite{OVCHINNIKOV2024114611}, the ordinary chaotic phase is preceded by the phase of the noise-induced chaos, where the TS is spontaneously broken by noise-induced antiinstantons ($\bar I$) and instantons ($I$) matching them. These are the processes of annihilation and creation of the pairs of solitons which are the left and right moving kinks. The solitons move at a fixed velocity, making the effective field theory (EFT) a Lorentian conformal field theory (CFT), despite the original SDE being Galilean. {\bf (right)} The AdS/CFT correspondence is an approximation where a CFT is described by a dual theory in the bulk of the AdS space with the conformal boundary being the spacetime of the original theory such as the one in the left. If the AdS/CFT correspondence is a reasonable approximation for the EFT of the BE, the dual theory is a cohomological TFT of AdS, making certain matrix elements, such as Wilson loops, natural objects of interest.}
\end{figure}

\subsection{TS Breaking and Generalization of Chaos}
\label{Sec_SUSY_Breaking}
Let us now recall (see, e.g., Ref. \cite{Torsten,OvcEntropy} and references therein) that most eigenstates of the supersymmetric evolution operator (SEO) are non-supersymmetric doublets, with their bras and kets given by
\begin{eqnarray}
    | i \rangle   \text{ and } \langle i | = \langle i' |\hat Q  \text{, and }   |i'\rangle = \hat Q | i \rangle \text{ and } \langle i' |,\label{NonSUSY_states}
\end{eqnarray}
with $\langle i| i\rangle=\langle i'| i'\rangle=\langle i'| \hat Q|   i\rangle $ , while the remaining eigenstates are supersymmetric singlets, $\theta$, such that
\begin{eqnarray}
    \langle \theta | [\hat Q, \hat X] | \theta \rangle = 0, \forall \hat X.\label{SUSY_requirement}
\end{eqnarray}
This property of supersymmetric eigenstates, combined with the fact that the SEO (\ref{SEOOper}) is $Q$-exact, implies that the eigenvalues of the supersymmetric eigenstates are exactly zero:
\begin{eqnarray}
    {   H}_\theta = \langle \theta | \hat H | \theta \rangle = 0.\label{ZeroEigenvalues}
\end{eqnarray}
The other eigenvalues, ${   H}_i = \langle i| \hat H|i\rangle$, are either real or appear in complex conjugate pairs, owing to the fact that the SEO is a real operator. This complex conjugate spectrum is a signature of yet another fundamental symmetry in STS, the pseudo-time-reversal or $\eta$T symmetry \cite{Mos021,Mos023}. 

The spontaneous breakdown of TS occurs when the ground state is non-supersymmetric. This situation arises when the model has eigenstates with negative real parts of their eigenvalues.\footnote{The possibility for the eigenvalues of evolution operator to have negative real parts is a distinctive feature of STS as compared to high-energy physics models.} In such cases, the eigenstates with the most negative real part of their eigenvalue dominate the long-time limit of the evolution, growing fastest. Consequently, the partition function grows exponentially in the long time limit:
\begin{eqnarray}
    Z|_{t\to\infty} = \sum_{i, \text{Re } {   H}_i = - \Lambda} e^{-{   H}_i t}\propto \text{Re } e^{-{   H}_0 t},\; \Lambda = -\min_i \text{Re }{   H}_i.\label{ExponentialGrowth}
\end{eqnarray}
where ${   H}_0= \langle 0| \hat H | 0\rangle$ is the eigenvalue of the ground state, which is one the fastest growing eigenstates and which will be used to define the generating functional in Eq.(\ref{generatingFunctional}) below.

The partition function represents the number of closed solutions and its exponential growth is a definitive feature of chaos in the dynamical system theory where the corresponding exponential growth rate, $\Lambda$, is identified as the dynamical entropy or dynamical "pressure" \cite{Rue02}. This proves that the spontaneous TS breaking in STS is the stochastic generalization of chaos of the dynamical systems theory.

\subsection{Deterministic vs. Noise-induced Chaos}
\label{Sec_SUSY_Breaking}

In STS, there are two major types of chaos. The first is conventional chaos (C-phase), where TS is broken by the non-integrability of the flow vector field in SDE. The second type is instantonic or noise-induced chaos (N-phase), where TS is spontaneously broken by noise-induced instantons. The N-phase can be though of as a C-phase boundary widened into a finite-width phase due to noise. It collapses onto the critical boundary of the C-phase in the deterministic limit.

The N-phase is a broad, multidisciplinary stochastic dynamical phenomenon commonly referred to in the literature as self-organized criticality (SOC) \cite{Bak87,A_recent_review_on_SOC,SOC_controvercy,Asc11,seizure}. The dynamics in the N-phase is dominated by (anti)instantonic processes such as earthquakes, solar flares, neuroavalanches, and many others. This dynamics can be informally described as a "solitonic liquid", where solitons appear/disappear in antiinstantonic/instantonic processes of creation/annihilation of tandems of solitons. One illustrative example is the overdamped stochastic sine-Gordon model \cite{OVCHINNIKOV2024114611}, where the N-phase dynamics consists of the left/right moving kinks and antikinks (see Fig. \ref{Fig_1}a). This model will serve as our primary example later in the paper.

To confirm that the TS is broken in this example, one can examine the model's partition function. This can be approximately calculated by recognizing that each multi-instanton configuration is uniquely determined by the spatiotemporal positions of its anti-instantons. Neglecting perturbative corrections, in line with the localization principle of supersymmetric models, the partition function becomes an integral over the moduli space of these multi-instanton configurations:
\begin{eqnarray}
    Z(T) \approx \sum_{i=0}^{\infty} \frac{1}{k!}\int_{0}^T dt \int dr e^{-\Delta U/\Theta} = e^{-T{   H}_0},
\end{eqnarray}
where the factorial $k!$ account for the fact that (anti)instatnons are identical (interchange of instantons lead to the same configuration), the exponential Gibbs factors come from the antiinstantons of the creation of kink-antikinks pairs with, $\Delta U$, being the "energy" barrier between the vaccum and the kink-antikink configuration, and the ground state eigenvalue is
\begin{eqnarray}
    {   H}_0 = -R e^{-\Delta U/\Theta},
\end{eqnarray}
where $R$ is the length of the spatial circle. As expected, the eigenvalue of the ground state has a negative real part and it is real so that the TS and pseudo-time-reversal symmetry are spontaneously broken and unbroken, respectively.

\subsection{The Meaning of Ghosts and the Butterfly Effect}
\label{Sec_Meaning_ofFermions}
To make good sense out of STS, one has to provide the physical interpretation of the Faddeev-Popov ghosts.\footnote{In the literature, ghosts are often thought to represent noise. This is a misconception which is particularly evident in cases where the noise-related term $g$ in Eq. (\ref{SDE}) is independent of $\varphi$—the so-called additive noise. In such scenarios, the ghosts are not even coupled to $g$. Instead, it is the Lagrange multiplier $B$ that represents the noise and is always coupled to $g$.} This can be done through the following observation. Recall that in numerical investigations of chaos, one typically defines a differential and propagates it along a trajectory to extract information about Lyapunov exponents. Multiple differentials can be propagated simultaneously. However, propagating two parallel differentials yields no additional information compared to propagating just one. This suggests that the mathematical representation of these differentials must be such that two parallel differentials do not make sense. This requirement is satisfied for anticommuting differentials that appear in differentials forms\footnote{For example, if $dx$ and $c dx$ is a pair of differentials pointing in the same direction with $c$ being some number, then their product $c dx\wedge dx = 0$.} -- the fundamental mathematical objects that are dual to volumes of different dimensionality. In field-theoretic terms, these anti-commuting differentials are Grassmann numbers or fermions just like the Faddeev-Popov ghosts of STS. Thus, the fermions in STS serve as a mathematical bookkeeping tool for the differentials that define Lyapunov exponents. 

In other words, while the traditional Fokker-Planck approach focuses on the time evolution of the total probability density, where the wavefunction represents an observer concerned solely with bosonic variables, STS adopts a different perspective. Here, the observer is a \emph{chaotician}, to borrow the term from the "Jurassic Park" blockbuster, who is interested not only in the trajectories of the bosonic variables but also in the evolution of the differentials. Accordingly, the wavefunction is not just a function, but a differential form.

From a more technical point of view, the Parisi-Sourlas approach can be viewed as a gauge-fixing procedure \cite{Baulieu_1988,Baulieu_1989,TFT_BOOK}. From this perspective, the ghosts are auxiliary fields needed to represent the functional determinant that appears on switching from the noise to the systems variables. This functional determinant is temporary nonlocal and in chaotic systems it hosts the BE. Thus, the ghosts of STS is a tool which allows to encode the temporarily nonlocal BE in the temporarily local wavefunction. 

\subsection{The Response and the Butterfly Effect}
\label{Sec_Generating_Functional}
The BE can be looked upon as a response of the system. To define it this way, let us consider perturbations of the SDE (\ref{SDE}) in the following general form:
\begin{eqnarray}
F \to F + \zeta_a(x) e^{a}(\varphi,x), \label{Perturbation}
\end{eqnarray}
where $\zeta$'s are external probing fields coupled to the system via a vector field, $e$. According to Eq.(\ref{Perturbation}), the pertrubed action is,
\begin{eqnarray}
    &S(\zeta) = S + [{\mathcal Q}, \int d^{d}x \zeta_{a}(x) v^{a}(\varphi,x)], \label{probing}
\end{eqnarray}
where $S$ from Eq.(\ref{Action_0}) and $v^{a}(\varphi,x) = i\bar\chi(x)e^{a}(\varphi,x)$. Eq,(\ref{probing}) shows that pertrurbations of the model on the level of the SDE leave the action $Q$-exact.

The next step is to come up with a suitable object, or a generating functional, that can be use to study the response of the model to probing fields. As mentioned in Sec. (\ref{Sec:STS}) and discussed in more details in \ref{WittenIndexButterflyEffect}, the Witten index cannot be used to define the response of the system. In the light of this, it may seem like a good idea to use the partition function (\ref{partitionFunction}) for this purpose. This would not be the best approach either because, in cases where the fastest-growing eigenstates are complex,\footnote{Such situation can be interpreted as a spontaneous breakdown of pseudo-time-reversal symmetry that all SDEs possess.} the partition function exhibits oscillatory behavior in the long-time limit (see Eq.(\ref{ExponentialGrowth})). A more practical alternative is to designate one of the few fastest-growing eigenstates\footnote{There are either two or four such fastest growing eigenstates depending on whether the ground state eigenvalue is real or complex, \emph{i.e.}, depending on whether the pseudo-time-reversal symmetry of the model is broken spontaenously.}  as the ground state, $|0\rangle$, and use it to define the following generating functional:
\begin{eqnarray}
    Z(\zeta) = \lim_{T\to\infty}\langle 0 | \hat M_{T/2, -T/2}(\zeta)| 0 \rangle,\label{generatingFunctional_eta}
\end{eqnarray}
where the perturbed SEO is given by Eq.(\ref{SEO}) with action (\ref{probing}).

The following correlators can be used to quantify the response of the system to perturbations of SDE,
\begin{eqnarray}
&\left.\frac{\delta}{\delta \zeta_{a_1}(x_1)}...\frac{\delta}{\delta \zeta_{a_k}(x_k)} Z(\zeta)\right|_{\zeta=0} \nonumber \\& = \langle 0| \mathcal{T} [\hat Q,\hat v^{a_1}(x_1)] ... [\hat Q,\hat v^{a_n}(x_n)]|0\rangle. \label{Rs}
\end{eqnarray}
where $\hat v(t) = \hat M_{-T/2,t}\hat v\hat M_{t,-T/2}$ and $\hat v$ are the perturbation operator (\ref{probing}) in the Heisenberg and Schroedinger representations respectively, and the TS operator, $\hat Q=\hat Q(t)$, is independent of time due to its commutativity with the SEO.

When TS is unbroken and the ground state is a supersymmetric singlet, all correlators (\ref{Rs}) vanish, as deduced from Eqs.(\ref{SUSY_requirement}) and the fact that the product of any number of $Q$-exact operators is also a $Q$-exact operator:
\begin{eqnarray}
    [\hat Q,\hat v^{a_1}(x_1)] ... [\hat Q,\hat v^{a_n}(x_n)] =  [\hat Q,\hat v^{a_1}(x_1) ... [\hat Q,\hat v^{a_n}(x_n)]].\label{Qexactnessofproduct}
\end{eqnarray}
Vanishing response (\ref{Rs}) can be interpreted as the absence of a system's response to perturbations in the limit of infinitely long evolution, the limit represented by a supersymmetric ground state. In other words, the system lacks long-term dynamical memory. 

Conversely, when TS is spontaneously broken, (\ref{Rs}) no longer vanish, indicating that the system responses to perturbations even in the limit of the inifnitely long evolution, represented by a non-supersymmetric ground state. This situation can be interpreted as the system's manifestation of the BE.

The perturbative analysis, such as the one used in this section, is a suitable tool for conventional perturbative phenomena where correlators decay over time -- such as in the case of 1/f noise (see Sec. \ref{Sec:OPE}). The BE, however, persists indefinitely. This exceptional longevity strongly suggests a topological origin for the BE. Further analysis within the EFT of the BE in Sec.(\ref{Sec:EFT}) provides additional support for this perspective.

\section{Global Symmetries and Background Supergauge Field}\label{Symmetries}

The TS (\ref{Eq_q}) is one of the global symmetries of the model. In the operator representation, it is defined by its nilpotent supercharge, 
\begin{eqnarray}
\hat Q \equiv \hat Q^+ =  \int \hat q(r) d^{d-1}r,\; \hat q(r) = \chi(r) i\hat B(r), \;\; (\hat{Q}^+)^2 =0.
\end{eqnarray}
Another global symmetry is the conservation of the number fermions. Its charge is,
\begin{eqnarray}
    \hat N = \int \hat n(r) d^{d-1} r,\; \hat n(r) = \chi(r) i\hat{\bar \chi}(r).
\end{eqnarray}
The supersymmetry of the Parisi-Sourlas approach is often called as $N=2$ supersymmetry because there is yet another global fermionic symmetry. Its nilpotent supercharge can be introduced as,
\begin{eqnarray}
\hat{Q}^- = \int  \hat{q}^-(x) d^{d-1}r, \; \; (\hat{Q}^-)^2 = 0.
\end{eqnarray}
This supersymmetry is a pseudo-time-reversal and pseudo-Hermitian counterpart of the TS such that
\begin{eqnarray}
   \hat H = [\hat Q^+, \hat Q^-],
\end{eqnarray}
which is Eq.(\ref{SEOOper}) with $\hat{\bar Q}$ from Eq.(\ref{Suerpcharges}) substituted by $\hat Q^-$. For Langevin SDEs \cite{TFT_BOOK}, all deterministic models including classical mechanics \cite{Gozzi_New}, the overdampped sine-Gordon model \cite{OVCHINNIKOV2024114611} (see Fig.\ref{Fig_1}a), and all other models where $\hat {\bar Q}$ is nilpotent, operator $\hat Q^-=\hat {\bar Q}$. In the most general case, $\hat {\bar Q}^2\ne0$ and operator $\hat Q^-\ne\hat {\bar Q}$, but $\hat Q^-$ can be constructed from the eigensystem of the SEO (see Eq.(141) of Ref.\cite{OvcEntropy}).\footnote{Explicit expression for this operator for a general class of SDEs has recently been established \cite{Weiderpass_Sethi}.} Its explicit form is not required for the further discussion, however. 

All the charges of the global symmetries commute with the SEO and form the $\mathfrak{gl}(1|1)$ superalgebra,
\begin{eqnarray}
    [\hat N, \hat Q^\pm] = \pm \hat Q^{\pm}, [\hat Q^+, \hat Q^-] = \hat H,\label{TheAlgebra}
\end{eqnarray}
where $\hat N$ and $\hat H$ are even elements, and $\hat Q^{\pm}$ are odd elements of the superalgebra. According to the definition of supercommutator, the first bracket in Eqs.(\ref{TheAlgebra}) denotes a commutator while the second denotes anticommutator. 

Following the framework of quantum field theory, we can couple the model to an external supergauge field, $A$, \cite{QUILLEN198589} by introducing the generating functional:
\begin{eqnarray}
    Z &\to& Z(A) = \lim_{T\to\infty}\langle 0 | \hat M_{T/2, -T/2}(A)| 0 \rangle,\label{generatingFunctional}
\end{eqnarray}
where the coupling is implemented through the use of covariant derivatives:
\begin{eqnarray}
    \partial_\mu \hat \Phi(x) &\to& \hat D_\mu \hat \Phi(x) = (\partial_\mu + \hat{A}_\mu(x))\hat \Phi(x),
\end{eqnarray}
with the supergauge field defined as:
\begin{eqnarray}
    \hat{A}_\mu(x) &=& A_\mu^\alpha(x) \hat r_{\alpha}(x).
\end{eqnarray}
Here, $\hat r = (\hat n, \hat q^+, \hat q^-, \hat h)$ represents the spatial densities of the corresponding charges, with index $\alpha$ running over these components. For example, $\hat H = \int \hat h(r)d^{d-1}r$, making $\hat h$ the time-time component of the energy-momentum tensor \footnote{The identification of $\hat h$ as the time-time component of the energy-momentum tensor implies that the corresponding symmetry is time translation. This observation suggests that, in certain cases, the symmetry group may be extended to include additional transformations of the base space. However, in the most general case of a nonhomogeneous base space, such an extended symmetry group does not exist.}. The densities satisfy the spatial extension of Eq.(\ref{TheAlgebra}), \emph{e.g.}, $[\hat n(r), \hat q^+(r')] = \delta^{d-1}(r-r')\hat q^+(r)$.

By a standard textbook argument (see, \emph{e.g.}, Ref.\cite{Seinberg}), a local transformation of the fields,
\begin{eqnarray}
    \hat \Phi(x) \to \hat g(x) \hat \Phi(x) \hat g^{-1}(x), \;\hat g(x) = e^{c^{\alpha}(x) \hat r_{\alpha}(x)},
\end{eqnarray}
where two of $c$'s are Grassmann even and two odd, in the generating functional (\ref{generatingFunctional}) can be compensated by a corresponding transformation of the background field,
\begin{eqnarray}
    \hat{A}_\mu(x) \to \hat{A}_\mu(x) - \hat g^{-1}(x)\partial_\mu \hat g(x),
\end{eqnarray}
so that the value of the generating functional remains unchanged. In other words, $Z(A)$ is supergauge invariant.

There is a subtlety in this argument that must be addressed. Specifically, it is necessary to verify that the path integration measure is also invariant under the local transformation of the fields. To check this, consider an infinitesimal transformation of variables:
\begin{eqnarray}
    \delta \hat \Phi(x) = c^\alpha [\hat r_\alpha(x), \hat \Phi(x)],
\end{eqnarray}
where $c$'s are small parameters. Under this transformation, the measure changes as follows:
\begin{eqnarray}
&\prod\nolimits_x D^4\Phi  \to \prod\nolimits_x \text{Ber}(1 + c^\alpha(x) \mathfrak{r}_\alpha(x)) D^4\Phi(x) \nonumber \\&= e^{ \int d^{d}x c^\alpha(x) \text{ sTr }\hat{\mathfrak{r}}_\alpha(x)} \prod\nolimits_x D^4\Phi(x), \label{measure}
\end{eqnarray}
where $D^4\Phi = \varphi DB D\bar\chi D\chi $ and $\text{sTr}$ and $\text{Ber}$ denote, respectively, the supertrace and the Beresian (or the superdeterminant) of linear operators acting in the $\Phi$-superspace \footnote{The integration over $x$ can be understood as the conventional trace over the space-time "indices"}. These linear operators are defined by $\mathfrak{r}$'s, which is the basis of the infinitesimal Jacobian of the variable transformation,
\begin{eqnarray}
    \mathfrak{r}_\alpha{}^i_j(x) = \partial \left([\hat r_\alpha(x), \hat \Phi^i(x)]\right)_{\hat \Phi \to \Phi}/ \partial \Phi^j(x),
\end{eqnarray}
where $\hat \Phi \to \Phi$ denotes switching back to the pathintegral picture and Latin indices run over the components of the superfield, $\Phi$. Two of these operators are model-independent:
\begin{eqnarray}
\hat{\mathfrak{r}}_{q^+} = \left(\begin{array}{cccc}
0&0&0&0\\
0&0&0&1\\
1&0&0&0\\
0&0&0&0
\end{array}\right), 
\hat{\mathfrak{r}}_{n} = \left(\begin{array}{cccc}
0&0&0&0\\
0&0&0&0\\
0&0&1&0\\
0&0&0&-1
\end{array}\right).
\end{eqnarray}
and both have vanishing supertraces. The other two operators are model-specific but their supertraces also vanish. This follows from the fact that $\hat{\mathfrak{r}}$'s is a representation of $\mathfrak{gl}(1|1)$ provided by fields $\Phi$, and according to Eq.(\ref{TheAlgebra}), $\hat{\mathfrak{r}}^{q^-}$ and $\hat{\mathfrak{r}}^{h}$ are supercommutators and supertraces of all supercommutators vanish. Consequently, none of $\hat{\mathfrak{r}}$ contribute to the exponent of Eq.(\ref{measure}). Thus the measure remains unchanged and the generating functional is indeed supergauge invariant.

\section{Effective Filed Theory}
\label{Sec:EFT}
The procedure for constructing EFT using the background field method is well established (see, \emph{e.g}, Ref. \cite{RFT_SSB_book}).  It begins with the introduction of the generating functional,
\begin{eqnarray}
    G(A)  &=& - \log \iint {\mathcal{D}}\bar\psi{\mathcal{D}}\psi e^{-S_\text{EFT}(A, \bar\psi,\psi )},\label{EFT_deinition}
\end{eqnarray}
which is expressed in terms of a collection of slow fields or order parameters, $\bar\psi,\psi$, that are relevant in the long-wavelength limit. 

These slow fields form irreducible representations of $\mathfrak{gl}(1|1)$, each of which is 2-dimensional \cite{GL1_1WZ_2005}. In other words,  the slow fields organize into supersymmetric doublets: $\psi_i = (\phi_i, \eta_i)$, where $\eta_i = \mathcal{Q}_\text{EFT}\phi_i, i=1,2,...$. The corresponding representations of $\mathfrak{gl}(1|1)$ are given by,
\begin{eqnarray} 
    \hat r_{q^+} = 
    \left(\begin{array}{cc}
    0 &  1 \\
    0 & 0
    \end{array}\right),\hat r_{q^-} = e_i \hat r_{q^+} ^T,\hat r_n = \hat \sigma_z /2 , \hat r_{h} = e_i \hat 1_{2\times2},
    \label{representation}
\end{eqnarray}
where $\sigma_z$ is the Pauli matrix, $e_i$ is the parameter of the representation that can be though of as a charge.

In addition, one must introduce the "anti-fundamentals" or momenta fields $\bar\psi = (\pi_i,\bar\eta_i), \mathcal{Q}_\text{EFT}\bar\eta_i=\pi_i$, which are direct analogues of $B$ and $\bar\chi$ in Sec.\ref{Sec:STS}. The overall TS now takes the form, 
\begin{eqnarray}
    \mathcal{Q}_\text{EFT} =\int d^{d}x\sum\nolimits_i ( \eta_i\delta/\delta\phi_i + \pi_i\delta/\delta \bar\eta_i).\label{TS_EFT}
\end{eqnarray}
The spontaneous breakdown of TS suggests the emergence of gapless, slow fields. But it does not imply that the original symmetry disappeared. The symmetry is still there but now it is realized in a way that involves transformations of all the slow fields. This means that the EFT must be $\mathcal{Q}$-exact\footnote{This is actually yet another way to see that the slow fields must come in supersymmetric doublets.} but not with respect to the original TS. Instead, it is $Q$-exact with respect to operator (\ref{TS_EFT}). For instance, the simplest version of the kinetic part of the Lagrangian density should consist of $Q$-exact terms like this, $\mathcal{Q}_\text{EFT} \bar\eta_i \hat L \phi_i = \pi_i \hat L \phi_i - \bar\eta_i \hat L \eta_i$, where $\hat L$ is d'Alembert or Dirac operators for Lorenzian or type A models (see Sec.(\ref{Lorenzian_Galilean}) below) and $\hat L = \partial_t - \partial_r^2$ for Galilean or type B models. 

Thus, the action in Eq.(\ref{EFT_deinition}), in the absence of the external supergauge field, can be given as:
\begin{eqnarray}
S_\text{EFT}(0, \bar\psi, \psi ) = \int d^{d}x L_\text{EFT}( ..., \partial_{\mu_1} ... \partial_{\mu_k} \psi_i,...),\label{EFT_Lagrangian_zero_A}
\end{eqnarray}
where the dots represent other slow fields or their derivatives\footnote{The space-time derivations act only on $\psi$ and not on anti-fundamentals, which can always be achieved by partial intergation.}, and
\begin{eqnarray}
    &L_\text{EFT}( ..., \partial_{\mu_1} ... \partial_{\mu_k} \psi_i,...) = [\mathcal{Q}_\text{EFT}, l_\text{EFT}(..., \partial_{\mu_1} ... \partial_{\mu_k} \psi_i,...)]\nonumber\\
    &= ... + [\mathcal{Q}_\text{EFT}, \partial_{\mu_1} ... \partial_{\mu_k} \psi_i] \frac{\partial l_\text{EFT}}{\partial (\partial_{\mu_1} ... \partial_{\mu_k} \psi_i)} + ...\nonumber\\
    &= ... + \hat r_+ \partial_{\mu_1} ... \partial_{\mu_k} \psi_i \frac{\partial l_\text{EFT}}{\partial (\partial_{\mu_1} ... \partial_{\mu_k} \psi_i)}  + ...\label{Q_exact_action}
\end{eqnarray}
with some functional $l_\text{EFT}$. Here we used that the bi-graded commutator with $\mathcal Q_\text{ETF}$ acts as a differentiation, and that
\begin{eqnarray}
    [\mathcal{Q}_\text{EFT}, \partial_{\mu_1} ... \partial_{\mu_k} \psi_i] = \partial_{\mu_1} ... \partial_{\mu_k} \hat r_+ \psi_i= \hat r_+ \partial_{\mu_1} ... \partial_{\mu_k}  \psi_i.\label{Q_action}
\end{eqnarray}

The next step is to covariantly couple the slow fields to the external supergauge field. The result is the following action:
\begin{eqnarray}\label{EFT_Lagrangian}
S_\text{EFT}(A, \bar\psi, \psi ) = \int d^{d}x L_\text{EFT}( ..., \hat D_{\mu_1} ... \hat D_{\mu_k} \psi_i,...),\label{EFTWithGauge}
\end{eqnarray}
where covariant derivatives are defined as
\begin{eqnarray}
\hat D_\mu = \partial_\mu + \hat A_\mu, \; \hat A_\mu = \hat r_\alpha A^\alpha_\mu.\label{DefOfhatA}
\end{eqnarray} 
The action (\ref{EFTWithGauge}) is no longer $Q$-exact. That is, the presence of $A$ explicitly breaks the TS. However, we can extend the definition of the TS to include transformation of the external supergauge field,
\begin{eqnarray}
    \mathcal{Q} = \mathcal{Q}_\text{EFT} + \mathcal{Q}_A,\label{TS_TOTAL}
\end{eqnarray}
such that
\begin{eqnarray}
[\mathcal{Q}_A, \hat D_\mu] = [\hat D_\mu, {\hat r}_{q^+}] = [\hat A_\mu, \hat r^+],\label{conditionGauge}
\end{eqnarray}
or 
\begin{eqnarray}
\mathcal{Q}_{A} = \int d^{d}x\sum\nolimits_{\mu} ( A_\mu^{n}\delta/\delta A_\mu^{q^+} + A_\mu^{q^-}\delta/\delta A_\mu^{h}).\label{TS_A}
\end{eqnarray}
With respect to this extended TS, action (\ref{EFTWithGauge}) is $Q$-exact,
\begin{eqnarray}
    &L_\text{EFT}( ..., \hat D_{\mu_1} ... \hat D_{\mu_k} \psi_i,...) = [\mathcal{Q}, l_\text{EFT}(..., \hat D_{\mu_1} ... \hat D_{\mu_k} \psi_i,...)]\nonumber\\
    &= ... + \hat r_+ \hat D_{\mu_1} ... \hat D_{\mu_k} \psi_i \frac{\partial l_\text{EFT}}{\partial (\hat D_{\mu_1} ... \hat D_{\mu_k} \psi_i)}  + ...\label{Q_exact_action_}
\end{eqnarray}
as follows from Eq.(\ref{Q_exact_action}) by replacing $\partial\to \hat D, \mathcal{Q}_\text{EFT}\to\mathcal{Q}$ and using the analogue of Eq.(\ref{Q_action}),
\begin{eqnarray}
    [\mathcal{Q}, \hat D_{\mu_1} ... \hat D_{\mu_k} \psi_i] = \hat r_+ \hat D_{\mu_1} ... \hat D_{\mu_k}  \psi_i,
\end{eqnarray}
which can be derived from Eq.(\ref{conditionGauge}) and $[\mathcal{Q}, \psi_i]= \hat r_+ \psi_i$.

The generating functional (\ref{EFT_deinition}) can now be understood as the result of partially integrating out fields in an "extended" EFT. This extended EFT, in addition to the slow fields, considers the external supergauge field as a fluctuating field. Its action,
\begin{eqnarray}
    S_\text{EFT}(A,\bar\psi,\psi) = [\mathcal{Q},\int l_\text{EFT}(..., \hat D_{\mu_1} ... \hat D_{\mu_k} \psi_i,...)d^{d}x],
\end{eqnarray}
is $Q$-exact, signifying that $\mathcal{Q}$ is the TS of the extended EFT. TS, in turn, is a resilient property that is difficult to eliminate through partial integration of fields. For instance, one form of partial integration is renormalization in its Wilsonian interpretation. The persistence of supersymmetry in such procedures in certain general classes of models is known and related with the no-anomaly and non-renormalization theorems \cite{SEIBERG1993469}.

In other words, it is reasonable to expect that the generating functional should not only be supergauge-invariant -- a standard feature of the background field method -- but also $\mathcal Q_{A}$-exact:
\begin{eqnarray}
     G(A) = [\mathcal{Q}_{A}, \omega(A)],\label{QExactGeneratingFunctional}
\end{eqnarray}
where $Z$ is given by Eq.(\ref{EFT_deinition}), and $\omega(A)$ is the associated gauge fermion. This implies that $W(A)$ takes the form of the action of another cohomological TFT, with the caveat that the action is strongly non-local, as discussed in Sec.\ref{Generating_functional}, where the $\mathcal Q_A$-exactness of the leading terms in $W(A)$ is demonstrated. In the context of the AdS/CFT correspondence addressed in Sec.\ref{Sec:BF_AdSCFT}, this nonlocality is effectively replaced by propagation in the emergent holographic dimension.


\section{EFTs as CFTs}
\label{Sec:CFT}
\subsection{Lorenzian vs. Galilean Models} 
\label{Lorenzian_Galilean}

The very concept of EFT can be described as a simplified theory in terms of fields relevant in the long-wavelength limit. In this limit, only gapless fields are relevant and their presence in chaotic models is guaranteed by the Goldstone theorem. 

The corresponding EFTs for the gapless fields are scale-invariant or conformal field theories (CFTs)\footnote{There is a subtle difference between scale-invariance and conformal invariance, but in most cases, these terms are interchangeable.}. These CFTs are supersymmetric due to the presence of TS\footnote{In the literature, supersymmetric CFTs are typically models based on supersymmetric extension of the conformal symmetry of the basespace. However, EFTs in STS have less structure as they are supersymmetric in a minimal sense and the TS is not part of the conformal symmetry of the basespace}. Moreover, due to the non-Hermiticity inherent in STS, some EFTs may correspond to non-unitary\cite{NonUnitaryCFTs} or complex \cite{ComplexCFTs,PhysRevLett.133.076504} CFTs. Furthermore, $\mathfrak{gl}(1|1)$ CFTs are believed to fall into the class of log-conformal CFTs \cite{GL1_1WZ_2005}. Although all of these classifications may turn out to be relevant in future work, the most important classification for our purposes is the following.

CFTs can be broadly divided into two major types: Lorentzian (or, upon Wick-rotation, Euclidean) and Galilean (see. \emph{e.g.}, Ref.\cite{Arjun_Bagchi_2009}).  A similar classification is found in the literature on EFTs (see \emph{e.g.}, Ref.\cite{RFT_SSB_book}), where Lorentzian and Galilean models are referred to as type A and type B, respectively. The structure of Eq.(\ref{SDE}) might suggest that type B models are most appropriate in the context of STS. This observation may be somewhat discouraging, as Lorentzian models are simpler from the technical point of view.

Fortunately, this is not always true; Galilean SDEs can yield Lorentzian EFTs. The model shown in Fig. \ref{Fig_1} provides a clear example of this situation. The key feature of its dynamics is the motion of free left- and right-moving kinks, traveling at constant velocities. As a result, the effective dynamics is Lorentzian with the following Lagrangian density:
\begin{eqnarray}
    &L_\text{D-sG,EFT}(A,\bar\psi,\psi) = [\mathcal{Q}, \sum\nolimits_{i=\pm} \bar\eta_{i}\hat D_i \phi_{i} + ...] \nonumber \\ &= \sum\nolimits_{i=\pm, x} \bar\psi_{i,x} (-1)^x\hat D_i \psi_{i,x} + [\mathcal{Q}, ...],\label{DsG_EFT}
\end{eqnarray}
where $\hat D_\pm = \hat D_t \pm \hat D_r$. The two pairs of supersymmetric fields introduced in Eqs. (\ref{representation}) and (\ref{TS_EFT}) represent the left- and right-moving kinks, with the index $x=0,1$ enumerating the supersymmetric partners $\psi_{i,0} = \phi_i,\psi_{i,1} = \eta_i$. The dots indicate interactions of (anti-)instantons with the solitons and themselves. The non-interacting part of this model can be identified as the $bc$ ghosts-superghosts model \cite{MARTINEC1988249,ALVAREZGAUME1988455,kausch1995curiositiesc2}.

As mentioned in Sec. \ref{Sec_SUSY_Breaking}, the N-phase, or self-organized criticality, is a widespread phenomenon. Its dynamics is dominated by the motion of solitons -- the boundaries of the instantons -- whose velocities can be assumed constant in many cases. As a result, there must exist a whole class of models where the ETFs are Lorentzian, even though the original SDEs are Galilean. For simplicity, the only speak of this class of models in this paper.

\subsection{Operator Product Expansion and 1/f Noise}
\label{Sec:OPE}
Scale-invariant models exhibit power-law correlations. In the context of TS breaking, these correlations must be understood as the 1/f noise -- the experimental signature of chaotic behavior in the form of the power-law correlations of various physical observables. Within the CFT framework, 1/f noise can be explained via the operator product expansion (OPE) -- the key feature of CFTs (see, \emph{e.g.}, Ref.\cite{ginsparg1988appliedconformalfieldtheory} and Refs therein).

The following arguments must be generalizable to a broader class of models because power-law correlators is a central feature of all CFTs. That said, we focus on the simplest case of CFTs: Euclidean CFTs, related to the corresponding Lorenzian CFTs by Wick rotation, in three or more dimensions with only scalar fields. In this context, there exists a set of primary fields, $\hat O_i, i=1,..,N$, each having descendants that can be symbolically given as: $\hat O_{i}^{(l)} = \partial_{\mu_1}...\partial_{\mu_l}\hat O_i, l\ge0$. The OPE says that a product of two operators can be given as a linear combination of the primaries and their descendants,
\begin{eqnarray}
    \hat{O}_i(x+y/2) \hat O_j (x-y/2) 
    =\frac{\delta_{ij}}{|y|^{2\Delta_i}} + \sum\nolimits_{k,l}C_{ijkl}(y)\hat O_{k}^{(l)}(x),\label{OPE}
\end{eqnarray}
where $k$ runs over the primary fields, $l$ runs over the descendants, and $C$'s are model-specific functions such that,
\begin{eqnarray}
C_{ijkl}(y) \propto 1/|y|^{\Delta_i+\Delta_j-\Delta_{k,l}}.\label{OPE_2}
\end{eqnarray}
Here, $\Delta$'s are conformal dimensions of the primary fields, $[\hat D, \hat O_i(x)] = \Delta_i \hat O_i(x)$, with $\hat D$ being the dilution operator of the CFT, and $\Delta_{k,l}=\Delta_k + l$ are conformal weights of the descendants.

The primaries are normally ordered operators so that $\langle0| \hat O_i^{(l)}(0) |0\rangle = 0$. Therefore,
\begin{eqnarray}
    \langle 0 | \hat O_i(0)\hat O_i(x)| 0 \rangle =1/|x|^{2\Delta_i}.
\end{eqnarray} 

The OPE is related with yet another fundamental concept in CFTs called the operator-state correspondence. It says that for any local operator, there is a corresponding state in the Hilbert space. Moreover, the states corresponding to the primary fields and descendants form a complete basis in the Hilbert space. Through the operator-state correspondence, the space of local operators in CFT borrows the structure of the Hilbert space. In particular, just like any state can be given as a linear combination of the basis vectors, any local operator can be given as a linear combination of the primaries and descendants.
\footnote{The OPE (\ref{OPE}) can be interpreted as this exchange of the structures between the Hilbert space and the space of operators, operating in reverse direction. Specifically, the two operators on the left-hand side of Eq. (\ref{OPE}) each create a state at the corresponding spacetime point. Just as with operators, the product of two states is also a state. The right-hand side thus represents the resolution of this composite state into the Hilbert space basis at the middle spacetime point.}

Accordingly, any physical observable that can be represented by a local normally ordered operator, $\hat O(x)$, so that $\langle0| \hat O(x) |0\rangle = 0$, can be expressed as,
\begin{eqnarray}
    \hat O(x) = \sum\nolimits_{i,l} c_{i,l}\hat O_i^{(l)}(x).\label{PrimaryDecomposition}
\end{eqnarray}
Its correlation function,
\begin{eqnarray}
   \left.\langle 0 | \hat O(0) \hat O(x)|0\rangle\right|_{|x|\to\infty} = c_{m,o}^2/|x|^{2\Delta_{\text{m}}} + ..., \label{Mellin}
\end{eqnarray}
where $\Delta_{\text{m}}$ is the smallest conformal weight, $c_{m,0}$ is the corresponding coefficient in Eq.(\ref{PrimaryDecomposition}),\footnote{Situations when $c_{m,0}=0$ can happen only accidentally so that Eq.(\ref{Mellin}) is valid in most cases.} and dots denote subleading terms that can be dropped in the long-wavelength limit. Eq.(\ref{Mellin}) can be interpreted as 1/f noise.

\section{Generating Functional}
\label{Generating_functional}

Once the EFT is established, standard field-theoretic methods can be used to evaluate the generating functional. While the functional is model-specific, its leading term is universal across a wide class of models and can be derived by considering the limit of a small background field where the coupling of the background field to the system is given by, 
\begin{eqnarray}
    S_\text{eff}(A,\phi) \approx S_\text{eff}(0,\phi) + \int d^{d+1}x A^\alpha_\mu(x) J^\mu_\alpha(x),\label{Perturbation}
\end{eqnarray}
where $J$'s are the corresponding conserved currents.

In the linear response manner, the generating functional (\ref{QExactGeneratingFunctional}) can be expressed as,
\begin{eqnarray}
    G({A}) = \int d^{d+1}x d^{d+1}x' A^\alpha_\mu(x) A^\beta_\nu(x') \langle  J^\mu_\alpha(x) J^\nu_\beta(x') \rangle/2 + ...\label{WGaussianTerm}
\end{eqnarray}
and the current-current correlator is (see, \emph{e.g.}, Refs.\cite{RevModPhys.91.015002} and \cite{GaugeFieldsInAdSCFT}): 
\begin{eqnarray}
\langle  J^\mu_\alpha(0) J^\nu_\beta(x) \rangle = \tau_{\alpha\beta} (g^{\mu\nu}\partial^2 - \partial^\mu\partial^\nu) f(x),\label{CurrentCurrent}
\end{eqnarray}
where $f(x)\propto x^{-2(d-2)}$, $g$ is the metric, and $\tau_{\alpha\beta}$ are model-specific coefficients. This form of current-current correlator follows from the conservation of the currents associated with the global symmetries, as dictated by Noether's theorem and Ward identities.

To explore the coefficients $\tau$'s, let us revisit the example in Fig.\ref{Fig_1}. In the conformally equivalent Euclidean formulation and radial quantization of Eq.(\ref{DsG_EFT}) -- where the complex plane serves as the basespace with coordinates $z,\bar z=x\pm iy$ -- the currents can be expressed as:
\begin{eqnarray}
J^{z}_\alpha = :\bar \psi_{+,w} r_\alpha^{wx}\psi_{+,x}:, \;  J^{\bar z}_\alpha = :\bar \psi_{-,y} r_\alpha^{yz}\psi_{-,z}:.
\end{eqnarray}
The fundamental two-point functions are $\langle \pi_{\pm} (0)\phi_{\pm}(z) \rangle=\langle \bar\eta_{\pm}(0) \eta_{\pm}(z) \rangle= G_\pm(z)=1/z^\pm, z^\pm \equiv z, \bar z$. Applying Wick's theorem and neglecting interactions, the correlators are:
\begin{eqnarray}
    \langle J^z_\alpha(0) J^{z}_\beta (z) \rangle = r_\alpha^{yz} r_\beta^{wx} \langle \bar \phi_{+,y}(z) \phi_{+,z}(0) \bar \phi_{+,w}(0) \phi_{+,x}(z) \rangle \nonumber\\=
    r_\alpha^{yz} r_\beta^{wx} \delta_{xy}\delta_{wz} (-1)^{\delta_{1x} \delta_{1y}} G_+(z)G_+(-z) = \text{sTr }\hat r_\alpha\hat r_\beta / z^2,\label{JJ_corr}
\end{eqnarray}
and similarly, $\langle J^{\bar z}_\alpha(0) J^{\bar z}_\beta (\bar z) \rangle = \text{sTr }\hat r_\alpha\hat r_\beta /\bar z^2$, while the remaining components vanish.\footnote{In Eq.(\ref{JJ_corr}), the sign feature of the supertrace, $(-1)^{\delta_{1x} \delta_{1y}}$, comes form the ordering of fermion operators.} Given the metric in $z,\bar z$ coordinates, $g_{\mu\nu} = (1/2)\hat \sigma_x, g^{\mu\nu} = 2\hat \sigma_x$, where $\sigma_x$ is the first Pauli matrix, it is evident that the current-current correlator is consistent with Eqs.(\ref{CurrentCurrent}), with $f(z)=\log z\bar z$ and coefficients given by,
\begin{eqnarray}
    \tau_{\alpha\beta} = \text{sTr }\hat r_\alpha \hat r_\beta. \label{constants}
\end{eqnarray}
From the mathematical point of view, Eq.(\ref{constants}) is the most natural choice, corresponding to the Casimir operator for $\mathfrak{gl}(1|1)$ (see below) and leading to a supergauge invariant generating functional. Moreover, the fact that $\tau$'s are traces of the pairs of generators is well-established in the theory of spontaneous breakdown of bosonic symmetries \cite{RFT_SSB_book}, and Eq.(\ref{constants}) is precisely that, with the adjustment that the trace is supersymmetric. We conclude that Eq.(\ref{constants}) is valid for a wide class of models. 

Returning to the generating functional and recalling that it must be gauge-invariant, we generalize Eq. (\ref{WGaussianTerm}) as:
\begin{eqnarray}
    G(A) &=& \int L_\text{YM}(x)d^{d}x + ...
\end{eqnarray}
where the Yang-Mills Lagrangian density is given by
\begin{eqnarray}
    &L_\text{YM}(x) = \text{sTr }\hat F_{\mu\nu} \circ \hat F^{\mu\nu}(x)\equiv\nonumber \\
    &\equiv \text{sTr }\int \hat F_{\mu\nu} (x+y/2) f(y) \hat F^{\mu\nu}(x-y/2) d^{d}y,
    \label{guageInvariance}
\end{eqnarray}
and the antisymmetric curvature 2-form is given by
\begin{eqnarray}
    \hat{F}_{\mu\nu} = [\hat D_\mu, \hat D_\nu],
\end{eqnarray}
with the following components,
\begin{subequations}
\begin{eqnarray}
    F^n_{\mu\nu} &=& \partial_{[\mu}  A^n_{\nu]},\\
    F^{\pm}_{\mu\nu} &=& \partial_{[\mu}  A^\pm_{\nu]} \pm A^\pm_{[\mu} A^n_{n]},\\
    F^{h}_{\mu\nu} &=& \partial_{[\mu}  A^h_{\nu]} + A^+_{[\mu}  A^-_{n]},
\end{eqnarray}    
\end{subequations}
where the square brackets in the subscripts indicate bi-graded antisymmetrization of indices, \emph{i.e.}, symmetrization if both factors are Grassmann odd and antisymmetrization otherwise. Eq.(\ref{guageInvariance}) generalizes Eq.(\ref{WGaussianTerm}) beyond the linear response approximation because the supergroup is non-commutative, and $F$ depend on the supergauge fields nonlinearly.

The supertrace in Eq.(\ref{guageInvariance}) pairs up components of $F$ according to the Casimir operator of $\mathfrak{gl}(1|1)$:\cite{GL1_1WZ_2005}
\begin{eqnarray}
    L_\text{YM} = e\;({F}^n_{\mu\nu} \circ {F}^{h,\mu\nu} - {F}^+_{\mu\nu}\circ {F}^{-,\mu\nu}),\label{YMLagrangian}
\end{eqnarray}
where $e=\sum_i e_i$. Using Eq. (\ref{TS_A}), we have $[\mathcal{Q}_{A}, {F}^+] = {F}^n$, $[\mathcal{Q}_{A}, {F}^h] = {F}^-$, and $[\mathcal{Q}_{A}, {F}^{n,-}] = 0$, so that, 
\begin{eqnarray}
    L_\text{YM} = [\mathcal{Q}_{A}, l_{YM}],\; l_\text{YM} = e {F}^+_{\mu\nu} \circ {F}^{h,\mu\nu}.\label{QexactYM}
\end{eqnarray}
This result is consistent with the discussion at the end of Sec. (\ref{Sec:EFT}), where it was argued that the entire generating functional should be $\mathcal{Q}_{A}$-exact.

From Eq. (\ref{TS_A}) and (\ref{conditionGauge}), it is clear that Eq. (\ref{QexactYM}) arises due to the Casimir operator of ${\mathfrak gl}(1|1)$ being a commutator, $\hat C = \hat r_n \hat r_h - \hat r_+ \hat r_- = [\hat r_+, \hat r_n \hat r_- ]$. This can be interpreted as $\hat C$ being $Q$-exact, with $\hat r_+$ playing the role of the TS. This property must be characteristic of a certain class of superalgebras, which evidently includes $\mathfrak{gl}(1|1)$.

\section{EFT and AdS/CFT correspondence}
\label{Sec:BF_AdSCFT}

\subsection{Applicability of AdS/CFT correspondence}

The theory of CFTs is well-established. Many important results have been obtained within it such as the current-current correlator in Eq.(\ref{CurrentCurrent}). Therefore, one possible way to proceed with the EFT of the BE is to employ the methods of CFTs that will certainly lead to interesting results. 

However, at this early stage in the development of the EFT for the BE, an approximate framework that correctly represents the situation on a qualitative level may be a more practical or desirable goal.
\footnote{Recall, for instance, that  the Ginzburg-Landau theory with only a few empirical parameters, provides a remarkably accurate description for, say, most superconductors, irrespective of their lattice structure or the specifics of electron-phonon interactions, among other details.}
It can provide initial insights and help build intuition, serving as a guiding tool for the subsequent formulation of a more rigorous and comprehensive theory of the BE. One such framework is the Anti-de-Sitter (AdS)/CFT correspondence \cite{AdSCFT_Maldacena,GUBSER1998105,Witten1998AntideSS,natsuume2016adscftdualityuserguide,Penedones_2016,KaplanAdSCFT}, which provides a dual "classical" description, where the conformal symmetry of CFT is realized as an isometry group of the plus-one dimensional AdS. 

AdS/CFT correspondence is an approximation. Even in high-energy physics models, where it was first discovered, it is believed to be exact only in certain limits, such as the large-$N$ limit. This approximation has been adopted in condensed matter physics, where it is thought to be relevant for a wide range of scale-invariant systems, such as phase transitions and critical phenomena. The emphasis here is less on exact results and more on qualitative descriptions of large-scale collective behavior. In this respect, the AdS/CFT meets the concept of EFT.

In fact, this positive assessment of the potential applicability of the AdS/CFT correspondence to EFT in STS can be made even more optimistic. It is essential to recognize that EFT itself is an approximation, primarily expected to yield qualitative results only at the tree level.
\footnote{This is one of the reasons why EFT is often understood as a classical field theory. For instance, the most interesting applications of the Ginzburg-Landau functional are such qualitative classical-field phenomena as Josephson current, Abrikov vortices, and Andreev refection in superconductors or modeling magnetodynamics in nanometer scale devices.} 
Within this context, the traditional question -- how well the holographic dual captures higher-order interactions in the original CFT -- becomes less crucial.

Instead, the AdS/CFT dual of an EFT must not necessarily be seen as an approximation layered on top of another approximation. Rather, it should be viewed as an alternative EFT. The central question, therefore, is not how precisely the dual EFT reproduces the original EFT, but rather which of the two -- the EFT or its AdS/CFT dual -- offers a more accurate, elegant, or otherwise beneficial description of the original model. In many instances, the AdS/CFT dual may present a more compelling option. Consequently, the AdS/CFT EFT for chaotic dynamics may be a fruitful investigation direction in its own right. This is the point of view we adopt here.

\subsection{The simplest AdS/CFT dual of EFT}

The procedure for constructing the AdS/CFT dual to a given EFT is well-established and extensively covered in textbooks on the subject \cite{natsuume2016adscftdualityuserguide,Penedones_2016,KaplanAdSCFT}. The dual may have varying field content depending on the desired level of detail: it may have a dynamic gravitational background or matter fields.\footnote{In the specific case of the model illustrated in Fig.~\ref{Fig_1}, the matter fields would correspond to two-component fermions and their superpartners, representing the left- and right-movers of the 2D EFT. The leading term of their action should be governed by the Dirac operator, covariantly coupled to $\mathcal{A}$ in the curved AdS background.} In its simplest setting, it contains only the supergauge field and can be expressed as the following GKP-Witten relation for the generating functional: 
\begin{eqnarray}
    {G}({A}) = \min_{\mathcal{A}}\left(\left.\mathcal{G}(\mathcal{A})\right|_{\mathcal{A}|_{B} = A}\right).\label{AdSCFT}
\end{eqnarray}
where $B$ stands for the conformal boundary of AdS. Here and in the following curly letters like $\mathcal A$ denote objects on the AdS side. The boundary condition signifies that the minimization must be performed under the constraint that the supergauge field on $B$ equals (up to some function of scale, see below) the external background field, $A$.    

The action in Eq.(\ref{AdSCFT}) has the following form,
\begin{eqnarray}
    \mathcal{G} = \int_\text{AdS} \left(c_{YM} \mathcal{L}_\text{YM} + c_{CS} \mathcal{L}_\text{CS}+...\right),\label{AdSAction}
\end{eqnarray}
where dots denote higher order terms. Constants $c$'s are model-specific strengths of the AdS versions of the Yang-Mills and Cherm-Simons (see below) term, which, unlike Lagrangian densities in the previous section, are local functionals of the supergauge field: $\mathcal{L}_\text{YM}(x) = \text{sTr } \hat {\mathcal F}\wedge \star (\hat {\mathcal F})$ where $\hat {\mathcal F}$ is the corresponding curvature 2-form and $\star$ is the Hodge star.

Eq.(\ref{AdSAction}) also explicitly includes the Chern-Simons term, yet another fundamental gauge-invariant object: in 3d, for example, ${L}_\text{CS,3d} = \text{sTr } \hat {\mathcal A} \wedge  \hat {\mathcal F}'$ where $\hat{\mathcal A}$ is the $\mathfrak{gl}(1|1)$ connection 1-form on AdS and $\hat {\mathcal F}'=d \hat{\mathcal A} + (2/3)\hat{\mathcal A} \wedge \hat{\mathcal A}$. In quantum field theory, Chern-Simon term may appear via parity anomaly \cite{Eduardo}. In STS, it may be the result of non-unitary of stochastic evolution. It may be expected to appear in situations with the spontaneously broken (pseudo-)time reversal or $\eta$T-symmetry (see Sec.\ref{Sec_Generating_Functional}). The breakdown of the $\eta$T-symmetry in the STS picture to the kinematic dynamo \cite{Torsten} is associated with the rotation of the galactic magnetic field. This observation along with the very mathematical meaning of the Chern-Simons term suggests that in case of AdS/CFT dual of EFT of BE this term may represent rotational aspects of dynamics, such as vorticity in turbulence. 

\subsection{The linear response}

The bulk gauge field at the boundary of AdS is the representative of the CFT currents. This field contains redundant degrees of freedom that can be eliminated by picking a certain gauge. The following gauge and coordinates are most convenient for a non-abelian gauge field in the context of the AdS/CFT correspondence \cite{Sundrum_AdSCFT_2012, KaplanAdSCFT}. This gauge choice is $\mathcal A_z=0$, where $z$ is the emergent "scale" dimension in the (Euclidean) Poincaré patch coordinates, $ds^2 = z^{-2}(dz^2 - g_{ij}dx^i dx^j )$. In this gauge, the number of components of $\mathcal A$ is the same as that of the gauge field in the CFT. To emphasize this fact, we introduce notation $\mathcal{A}_\mu = (0, \mathcal{A}_{\bar\mu})$, where the bar over the subscript means it runs over $t$ and $r$, but not $z$.

The above gauge leaves a residual gauge invariance with respect to gauge transformations independent of $z$. This residual gauge freedom can be further fixed by imposing the condition $\partial^{\mu} \mathcal{A}_{\mu} = \partial^{\bar \mu} \mathcal{A}_{\bar \mu} = 0$.

To investigate the linear response of the system, we consider the case of Yang-Mills term only, rescale the supergauge fields so that $c_{YM}=1$, and assume the weak coupling limit, in which the linear response makes good sense. The analysis goes along the traditional lines of the AdS/CFT correspondence with the only difference with the traditional case of bosonic symmetries (see, e.g., \cite{GaugeFieldsInAdSCFT}) being the supersymmetric nature of the gauge field. Particularly, in the gauge discussed above, the gaussian part of the action,
\begin{eqnarray}
    \mathcal{G} = \int dz d^{d} x \sqrt{\mathfrak{g}} \partial_{\mu} \mathcal{A}^a_{ \bar \nu} \mathcal{F}_a^{\mu \bar \nu},
\end{eqnarray}
where $\mathfrak{g}$ is metric on AdS, $\sqrt{\mathfrak{g}}=\sqrt{|\det \mathfrak{g}|}=z^{-(d+1)}$, and raising and lowering of the superalgebra assumed by $\hat \tau$ from Eq.(\ref{constants}), \emph{e.g.},  $\mathcal{F}_a = \tau_{ab} \mathcal{F}^{b}$. Integrating by parts,  
\begin{eqnarray}
    \mathcal{G} = \int d^{d} x \left.\mathcal{A}^a_{\bar \mu} \sqrt{\mathfrak{g}} \mathcal{F}_a^{z\bar \mu}\right|_{z\to 0} - \int dz d^{d} x \mathcal{A}^a_{\bar \mu} \partial_{\nu} \sqrt{\mathfrak{g}} \mathcal{F}_a^{\bar \mu \nu}.\label{curlyW_0}
\end{eqnarray}
The second, bulk term vanishes due to the equations of motion,
\begin{eqnarray*}
    \delta \mathcal{G}/ \delta \mathcal{A}^a_{\bar\mu} = 0 \to \partial_{\nu} \sqrt{\mathfrak{g}} \mathcal{F}_a^{\nu \bar\mu} = 0,
\end{eqnarray*}
or, using $\mathcal{F}_a^{\nu \bar\mu} = \tau_{ab}\mathfrak{g}^{\nu\eta}g^{\bar\mu \bar \xi} \partial_{\eta} \mathcal{A}_{\bar\xi}^b$ and turning to the momentum space for $x$,
\begin{eqnarray}
    ( z^{d-3}\partial_{z} z^{-(d-3)} \partial_{z} + k^2 ) \mathcal{A}^a_{\bar \mu}(z,k) = 0.
\end{eqnarray}
The solution can be given via Bessel functions, $\mathcal{A}(z,k)=c_1 z^{\Delta/2} J_{\Delta/2} (k z) +  c_2 z^{\Delta/2} Y_{\Delta/2} (k z), \Delta = d-2$. One constant is removed by the wave boundary at the bulk of AdS, $z\to\infty$, which turns the solution into the Hankel function of the first kind. The remaining constant is established by the boundary condition at the conformal boundary, $\mathcal{A}(k)|_{z\to0} = A(k)$, leading to $\mathcal{A}(z,k)|_{z\to0} = A(k) + z^{d-2} r(k) A(k)$, where $r(k)\propto k^{\Delta}$. 

Following AdS/CFT dictionary, \cite{AdSCFT_book} the slow falloff term must be identified as the extrnal guage fieeld on the CFT side, $\mathcal{A}(0) = A(k)$, and the fast falloff term  is the response current, $J_a^{\bar \mu}(k) = \sqrt{\mathfrak{g}} \mathcal{F}_a^{z\bar \mu}(k) = \tau_{ab} z^{-(d-3)} \partial_z \mathcal{A}^b_{\bar\mu}(z,k)_{z\to0} \propto \tau_{ab} r(k) A^b_{\bar\mu}(k) $. Eq.(\ref{curlyW_0}) reduces to  Eq.(\ref{WGaussianTerm}) with $\tau_{ab}r(k)$ being the AdS/CFT version of the current-current correlator. In the direct space, $r(x)\propto 1/x^{2 (d-1)}$ in accordance with Eq.(\ref{CurrentCurrent}). 

\subsection{The TFT aspect of the AdS/CFT dual of the EFT}

As discussed in Sec. \ref{Sec:OPE}, the 1/f noise can be attributed to the conformal symmetry of the effective field theory (EFT), realized as the isometry of the basespace within the AdS/CFT correspondence framework. While 1/f noise is a long-range phenomenon, it is not as enduring as the butterfly effect, which persists indefinitely and cannot be fully explained by the scale invariance of the EFT. Instead, the butterfly effect must be fundamentally tied to the other qualitative feature of the EFT in STS: the TS, which is also present on the AdS side of the AdS/CFT correspondence.

Namely, both terms in Eq.(\ref{AdSAction}) are $Q_\mathcal{A}$-exact as can be shown in direct analogy with Eq.(\ref{QexactYM}), \emph{e.g.}, ${L}_\text{CS,3d} = [\mathcal{Q}_{A}, {\mathcal A}^+ \wedge ({\mathcal F}')^{h}]$. Recalling also the discussion in Sec.\ref{Sec:EFT}, one can expect that the action of AdS/CFT dual EFT is $\mathcal Q$-exact in a general class of models including the ones considered here, 
\begin{eqnarray}
    \mathcal{G}(\mathcal{A}) = [\mathcal{Q}_{\mathcal A},\omega({\mathcal A})],\label{MoreGeneralAction}
\end{eqnarray}
where $\omega({\mathcal A})$ is yet another gauge fermion.

As a classical field theory, Eq.(\ref{MoreGeneralAction}) presents a subtle challenge: it may not be easy to assign a clear interpretation to the Grassmann-odd components of $\mathcal A$. This issue is not present in the more realistic version of AdS/CFT dual of EFT where the supergauge field only serves as a virtual interaction between the matter fields coupled to the $\mathcal Q$-exact boundary sources representing the probing fields in Eq.(\ref{probing}). Addressing such theories is beyond the scope of this paper. To complete our discussion, however, we bypass this issue by relaxing the boundary condition for the supergauge field and allowing $\mathcal A$ to fluctuate. This turns the model into a fully-fledged quantum field theory,
\begin{eqnarray}
    \mathcal{Z} = \iint e^{-[\mathcal{Q}_{\mathcal A},\omega({\mathcal A})]} \mathcal D\mathcal{A}.\label{dualTFT}
\end{eqnarray}
In general, introduction of fluctuations can lead to qualitative changes in a theory. In our case, however, the model is a cohomological TFT due to $\mathcal{Q}_{\mathcal A}$-exact action. Such theories, along with other supersymmetric models, exhibit a property known as the localization principle, according to which the pathintegration reduces to contributions from classical solutions because fluctuations from supersymmetric partners cancel each other out. Consequently, the extension (\ref{dualTFT}) of the AdS/CFT dual of EFT may be acceptible at least in some cases.

Cohomological TFTs offer a possibility to compute topological invariants in the following form,\footnote{There are two ways to understand this expression: as a matrix element on the global supersymmetric ground state or as an instantonic matrix elements between perturbative supersymmetric states.}
\begin{eqnarray}
\langle\mathcal{O}_{i_1} ... \mathcal{O}_{i_k}\rangle = \iint \mathcal{O}_{i_1} ... \mathcal{O}_{i_k} e^{-[\mathcal{Q}_{\mathcal A},\omega({\mathcal A})]}\mathcal{D} \mathcal{A},
\label{TopInariants}
\end{eqnarray}
where $O_i$ is a set of superguage invariant operators which are $\mathcal Q_\mathcal{A}$-closed, \emph{i.e.}, $[\mathcal{Q}_{\mathcal A},\mathcal{O}_i] = 0$, but are not $\mathcal Q_\mathcal{A}$-exact, ${\mathcal O}_{i} \ne [{\mathcal Q}_{\mathcal A}, o], $ $\forall i, o$. 

In the context of STS, matrix elements (\ref{TopInariants}) are interesting because they may help quantify the spatiotemporal structure of the BE. For this purpose, Wilson loops (see Fig.\ref{Fig_1}b) may be useful, which in our case are defined as,
\begin{eqnarray}
    \mathcal{O}_{C} = \text{sTr } {\mathcal T}_C e^{\int_C \hat {\mathcal A}_\mu(x) dx^\mu},\label{WilsonLoops}
\end{eqnarray}
where $C$ denotes a closed contour in the basespace, and ${\mathcal T}_C$ represents path ordering along $C$. Using Eq.(\ref{conditionGauge}) and recognizing that commutation with $\mathcal Q_{\mathcal A}$ acts as a differentiation, it follows that $[\mathcal{Q}_{\mathcal A},\Omega(\hat{\mathcal A)}] = [\Omega(\hat{\mathcal A)},\hat r^+]$ for any function $\Omega$. Thus:
\begin{eqnarray}
    [\mathcal{Q}_{\mathcal A},\mathcal{O}_{C}] = \text{sTr } [{\mathcal T}_{C} e^{\int_C \hat {\mathcal A}_\mu(x) dx^\mu}, \hat r^+] = 0,
\end{eqnarray}
because the supertrace of a supercommutator is zero. This demonstrates that Wilson loops are $\mathcal Q_\mathcal{A}$-closed. At the same time, it is not immediately evident that Eq.(\ref{WilsonLoops}) are always $\mathcal Q_\mathcal{A}$-exact. This suggests that Wilson loops could be suitable candidates for the operators $O$ in Eq.(\ref{TopInariants}). Even if this turns out not to be the case, there are other classes of operators in TFTs (see, \emph{e.g.}, \cite{TFT_BOOK}) that may fulfill the role of $O$'s in Eq.(\ref{TopInariants}). 

The evaluation of the matrix elements in Eq.(\ref{TopInariants}) may be challenging from a technical standpoint. However, it is hoped that some results on Wilson loops involving superconnections in high-energy physics models (see, e.g., Refs.\cite{WilsonLoopsMaldacena,WilsonLoops1} and refs therein) can be adopted into the EFT of the BE. In addition to these technical difficulties, an equally critical aspect of future work lies in establishing a compelling interpretation for objects like Eq.~(\ref{TopInariants}) and connecting them to the traditional understanding of chaotic dynamics.

\section{Conclusion}\label{conclusion} 
Within the supersymmetric theory of stochastic dynamics, the stochastic generalization of dynamical chaos is the spontaneous breakdown of the topological supersymmetry and the corresponding effective field theory is essentially a consistent description of the butterfly effect. In this work, we proposed the covariant background field method as a tool to construct such theories. 

We discussed the key features of these effective field theories, which, as conformal field theories, account for the 1/f noise through the operator-product expansion and operator-state correspondence. We showed that the associated generating functionals also possess topological supersymmetry. This observation unveils the topological nature of the butterfly effect. As a result, in models where the AdS/CFT correspondence is an acceptable approximation, the dual holographic description of the BE is a cohomological topological field theory. In this perspective, the conformal isometry of AdS underpins the 1/f noise in chaotic dynamics, while the topological supersymmetry is the origin of the infinitely-long butterfly effect.


\section*{Acknowledgments}
The author acknowledges the initial support from DARPA BAA "Physical Intelligence" and extends special gratitude to Kang L. Wang for his pivotal role in enabling this work. Appreciation is also extended to Savdeep S. Sethi, Gabriel A. Weiderpass, Torsten A. Ensslin, Robert N. Schwartz, Massimiliano Di Ventra, Ben Israelii, Daniel Toker, Skirmantas Janusonis, Dmitri A. Riabtsev, Cheng-Zong Bai, and Eugene Ingerman, all of whom positively influenced this work.

\section*{Declaration of competing interest}
The author declares no competing interests.

\appendix

\section{Witten Index and Response}
\label{WittenIndexButterflyEffect}
In the Literature on Parisi-Sourlas approach, it can often be encountered that the pathintegral representation of the model with periodic boundary conditions for fermions, or Witten index, is used to calculate various averages. This approach is not entirely accurate, however. To see this, let us introduce the following perturbed version of the Witten index:
\begin{eqnarray}
    W(\zeta) = \lim_{T\to\infty} \text{Tr } (-1)^{\hat N}\hat M_{T/2, -T/2}(\zeta),\label{WittenIndex_A1}
\end{eqnarray}
which is the periodic boundary condition analogue of Eq.(\ref{generatingFunctional_eta}) or the perturbed version of Eq.(\ref{WittenTopInv}). Using Eqs.(\ref{NonSUSY_states}), (\ref{SUSY_requirement}), (\ref{Qexactnessofproduct}), and (\ref{WittenIndex_A1}), one has,
\begin{eqnarray}
    & \left.\frac{\delta }{\delta\zeta_{a_1}(x_1)}...\frac{\delta }{\delta\zeta_{a_k}(x_k)} W(\zeta)\right|_{\zeta=0}  = \nonumber \\
    &=-\sum_{i} (-1)^{N_{i}} \langle i' |[\hat Q, [\hat Q, \mathcal{T}\hat v^{a_1}(x_1) ... [\hat Q, \hat v^{a_k}(x_k)]]]| i \rangle  \nonumber \\ 
    &+ \sum_{\theta} (-1)^{N_\theta} \langle \theta | [\hat Q, \mathcal{T} \hat v^{a_1}(x_1) ... [\hat Q, \hat v^{a_k}(x_k)]] |\theta \rangle =0, \label{Witten_index_Invariance}
\end{eqnarray}
where $\mathcal{T}$ is chronological ordering and $N$ is the number of fermions is the corresponding eigenstate. The first term in Eq.(\ref{Witten_index_Invariance}) vanishes because of the nilpotency of TS: $\hat Q^2=0$ and $[\hat Q, [\hat Q, \hat X]]=0, \forall \hat X$. The second term vanishes due to Eq.(\ref{SUSY_requirement}).

Eq.(\ref{Witten_index_Invariance}) suggests that the Witten index or the pathintegral representation of the model with periodic boundary conditions for fermions is independent of the probing fields. It cannot be used to explore the response of the system to external perturbations. As already pointed out in Sec.\ref{Sec:STS}, this is a reflection of the fact that the Witten index is a representation of the partition function of the noise and not of the model, and the noise has no information about the dynamics in the model.

Eq.(\ref{Witten_index_Invariance}) results particularly from the $\mathcal Q$-exactness of the probing operators in Eq.(\ref{probing}). However, in the literature, the Witten index is more commonly used to calculate averages for a broader class of operators,
\begin{eqnarray}
    \iint X(\varphi)P(\xi) \mathcal{D}\xi \to  \iint_{P.B.C.} X(\varphi) e^{-S} \mathcal{D}\Phi,\label{hybrid}
\end{eqnarray}
where $X$ is some functional of the fields in the model. While such averages do not vanish, Eq. (\ref{hybrid}) is still an unsuitable approach. To see this, consider the simple case of $0+1$ Langevin SDE, $\dot \varphi = - \nabla U(\varphi) + (2\Theta)^{1/2} \xi$, with $U$ of Morse type, \emph{i.e.}, with isolated critical points, $c, \nabla U(\varphi_c) =0$. In the deterministic limit, only critical points contribute into the pathintegral and each contribution comes with a sign factor. For example, 
\begin{eqnarray}
    \iint_{P.B.C.} \varphi(t) e^{-S} \mathcal{D}\Phi = \sum\nolimits_{c} \varphi_{c}(-1)^{ind(c)}.\label{NoSense}
\end{eqnarray}
This expression is intended to represent the mean value of $\varphi$, but it fails to do so. In fact, it lacks any meaningful statistical interpretation. In conclusion, treating the Witten index as if it were a partition function is a conceptual mistake.

\bibliographystyle{elsarticle-num} 
\bibliography{main}

\end{document}